# Gridbot: An autonomous robot controlled by a Spiking Neural Network mimicking the brain's navigational system

Guangzhi Tang
Computational Brain Lab
Rutgers University
New Brunswick, New Jersey
gt235@cs.rutgers.edu

Konstantinos P. Michmizos
Computational Brain Lab
Rutgers University
New Brunswick, New Jersey
konstantinos.michmizos@cs.rutgers.edu

## ABSTRACT

It is true that the "best" neural network is not necessarily the one with the most "brain-like" behavior. Understanding biological intelligence, however, is a fundamental goal for several distinct disciplines. Translating our understanding of intelligence to machines is a fundamental problem in robotics. Propelled by new advancements in Neuroscience, we developed a spiking neural network (SNN) that draws from mounting experimental evidence that a number of individual neurons is associated with spatial navigation. By following the brain's structure, our model assumes no initial all-to-all connectivity, which could inhibit its translation to a neuromorphic hardware, and learns an uncharted territory by mapping its identified components into a limited number of neural representations, through spike-timing dependent plasticity (STDP). In our ongoing effort to employ a bioinspired SNN-controlled robot to real-world spatial mapping applications, we demonstrate here how an SNN may robustly control an autonomous robot in mapping and exploring an unknown environment, while compensating for its own intrinsic hardware imperfections, such as partial or total loss of visual input.

## 1 INTRODUCTION

Navigating in complex and dynamic environments is a crucial yet seamlessly "effortless" task for the human brain. Advanced as they may have become, robots cannot currently exhibit such navigational skills, especially seen under the criteria for robustness, adaptability and efficiency. A promising path towards duplicating a human-like behavior is to mimic its underlying neural activity. Whereas most research has employed non-spiking neural networks to reproduce behavior [2, 11, 58], there is now a large interest in investigating the capabilities of Spiking Neural Networks (SNNs) for these tasks [8, 9, 53, 60]. Recent developments of large-scale neuromorphic hardware offering unprecedented asynchronous parallelism and power efficiency, namely the Intel Loihi [12], the IBM TrueNorth [48], as well as the BrainScaleS [59] and the SpiNNaker [18], have further spurred the development of SNN-based robotic controllers. Interestingly enough, and despite the emerging interest on autonomous machines, studies on SNN-controlled robot navigation have been rather sparse and rather limited to giving a proof of concept [9, 28, 29]. While there is definitely value in studying simplified tasks and basic control representations, there is a growing need to propose new SNN-controlled autonomous systems capable of naturally handling more complex scenarios and fully exploiting the advantages of neuromorphic hardware.

Abstracting away computational principles from the brain architecture, neuromorphic hardware promises relatively fast and power-efficient computations that can overcome the resource limitations of most mobile robots. However, to realize these promises, neuromorphic chips highly rely on the structure and computational principles of the underlying SNNs. With limited resources on asynchronous neuromorphic cores and local spike-time dependent plasticity (STDP) learning mechanisms, current neuromorphic systems can only leverage SNNs by keeping the number of synaptic connections and that of simultaneously active neurons as low as possible, contrary to the typical all-to-all initial connectivity found in the conventional neural networks. Interestingly, the most computationally expensive operation of a neuromorphic hardware is learning [45]. A different approach, which we propose here, is to imitate the brain as hierarchically organized layers of neural processing that minimizes the needs for learning. Once we replicate the brain's connectome, the targeted behavior emerges. Arguably, a few brain networks have been mapped at cellular resolution, challenged by the sheer size and volume of the vertebrate brain [44], the absence of a linkage between in-vitro cellular recordings and behavior [37], and the ineligibility of humans to become subjects in emerging in-vivo optical cellular imaging techniques [4, 23, 35]. Nonetheless, when constructed at this fundamental level, networks of brains at the lower end of the phylogenetic scale exhibit an unprecedented resolution in relating clusters of neural elements to clusters of behavioral phenotypes. For example, in the yet unique case of the nematode C.elegans whose neuronal wiring has been comprehensively mapped [66], neural networks following the brain structure have offered important insights for the behavioral contributions of small-world neural architectures [65], specific network elements [34], as well as the localized neural dynamics during the interaction with the environment [33].

For spatial navigation, the primate brain uses esoteric cues from the body and external environmental landmarks to locate itself, map its surroundings, and plan efficient routes towards its goals. Over the past decades, a large set of specialized neurons have been found to form what is now called the brain's navigational system [20]: Grid Cells (GC) in the medial entorhinal cortex (mEC) are related to speed integration and localization; Place Cells (PC) in the hippocampus are related to path integration, planning and memory; Border Cells (BC) represent environmental information; Head Direction Cells (HDC) are limbic neurons that provide orientation information. Goal Cells (GoalC) represent different goal locations. Despite the multitude of experimental studies, how the observed behavior emerges from the interconnectivity among the aforementioned and other cells remains a mystery. Therefore, any comprehensive bioinspired model employing these neurons needs to adhere to a number of extrapolations that will fill in the gaps of knowledge [52].

Here, we describe the results of our ongoing effort to develop a biologically constrained SNN of the brain's navigational system that will control the "Gridbot", an autonomously moving robot that we built in the lab (Figure 1). Specifically, we propose a basic connectome among the neural cells mentioned above, and neuromorphic algorithms that rely on nonlinear dendritic processes to allow neurons to connect to each other, synergize in either a static or plastic fashion, and combine cues from self-motion and the environment to represent locations, at their firing. This work suggests a neuromorphic navigation method for autonomous robots and it also proposes a connectome of distinct neurons from which the targeted behavior emerges.

## 2 RELATED WORK

A navigational system can be divided in 3 parts: simultaneous localization and mapping (SLAM), route planner and motor control. In computational neuroscience, self-localization mechanisms in navigation have been well studied, with two of the most popular approaches being the attractor and the oscillation model [3]. These models, however, assume static environments and do not consider the robust mechanisms related to the brain's navigational system. Goal-directed navigation uses a forward linear look-ahead probe to find goal direction cooperation with HDC, GC, PC and simulated prefrontal cortex cells (PFC), but it relies on assumptions for the PFC's topological relations which nevertheless are not represented in a neuromorphic way [16].

In robotics, SNNs are emerging as an approach for solving control problems in navigation and manipulation. A neuro-inspired SLAM method is RatSLAM [49]. Early versions of it can be implemented in a neuromorphic hardware by employing spatial neurons and local synaptic learning but they could not solve the data association and loop closure problems in realistic environments [50]. The current RatSLAM solves these problems by using an experience map, which does not have a neuromorphic analogy. There are route planners with results comparable to the classic A* planner [29]. SNNs are also used as motor controllers in insect robots [10] and robotic arms [14]. These SNNs can be seen as variations of a conventional linear controller that adapts to its environment without internally modeling it.

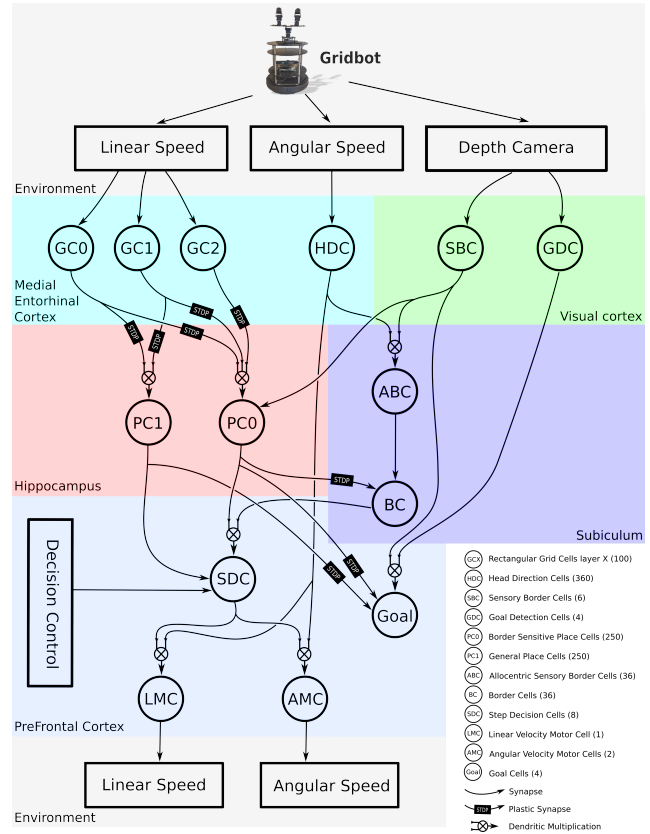

**Figure 1: The SNN-controlled Gridbot: The SNN structure follows the proposed connectome extrapollating biologically plausible connections among experimentally found neural cells that are associated with the brain's navigational system. The two main computational principles are depicted to process, and learn, dynamically neural representations of the surrounding environment: STDP and dendritic multiplication.**

## 3 A SPIKING NEURAL NETWORK OF THE BRAIN'S NAVIGATIONAL SYSTEM

We developed a biologically constrained network of 1321 spiking neurons (Figure 1). To construct our model, we incorporated: 1) Neurons that were modeled as Leaky Integrate-and-Fire (LIF) units [31]; 2) Synapses that were either hardwired or underwent plastic changes through STDP [51]; and 3) Dendritic trees that integrated synaptic inputs using nonlinear computations, including multiplication [46]. By following these biological constrains, our model had an intrinsic parallelism on information processing and learning.

Specifically, our proposed SNN encoded sensory information acquired from arbitrary environments into distributed maps and generated motor commands to control the robot movement. These maps were used for guiding the robot to search the environment without colliding into walls and they can also be used for planning efficient routes to goals in the near future. Learning took place at multiple neural layers: Border information in the camera-reference level, acquired from an RGB-Depth camera, was first transformed to world-reference space in BCs and eventually learned as synaptic weights on synapses from PCs to BCs (see Figure 1). Goal information was learned in a similar way. Self-motion information from odometers was initially integrated in rectangular GCs and eventually represented by PCs recruited from a pool of neurons with distinct single preferred place fields. The robot was controlled via the firing of specialized (motor) cells with the current goal being the exhaustive search of the environment without collisions.

## 3.1 Head Direction Cells and Rectangular Grid Cells

Self-motion information from odometer sensors, namely angular and linear speed, was integrated in mEC by HDCs and rectangular GCs using Continuous Attractor Network (CAN) models [7, 63]. The HDC layer consisted of 360 neurons, one for each rotational degree. Each HDC had a single preferred head direction for which it fired maximally. Direction coding from HDCs provided a world reference frame of the robot's heading that was independent of the robot's pose.

Rectangular GCs were neurons with multiple preferred place fields, emulating their biological counterparts. The preferred place fields of rectangular GCs formed a rectangular lattice instead of a triangular lattice for computational efficiency. There were 3 layers of rectangular GCs with different spacing between different preferred place fields of a single neuron, ranging from 25 cm to 1 m. Each layer consisted of 100 neurons forming a 10x10 grid that spanned the 2D space. Location coding from GCs provided a world-reference frame of the robot's location that was independent of the robot's pose.

## 3.2 Transformation from Camera-reference Space to World-reference Space

Figure 2a shows the transformation of border information from the camera-reference space (egocentric) to the world-reference space (allocentric). RGB-Depth camera information was transformed into a laser scan with distances and angular offsets of discrete obstacle points. Egocentric border information from the laser scan was first represented in Sensory Border Cells (SBCs). Without head direction information, SBCs could only have preferred receptive fields in the camera-reference space. Therefore, SBCs had receptive fields with distinct angular offsets on the robot's heading and distinct distances away from the border.

Border information without the allocentric direction between the border and the robot was not practical in controlling robotic movements. That is why egocentric border information in SBCs was transformed into allocentric border information in Allocentric sensory Border Cells (ABCs). To do so, we used dendritic multiplication to transform information in SBCs from the camera-reference space to the world-reference space on head direction represented by HDCs. Dendritic multiplication allowed a neuron to be activated only if all synapses on its dendritic tree have strong enough current inputs [46], similarly to a logical AND operation. In the proposed SNN, the ABC layer consisted of 360 neurons, the same number of neurons that the HDC layer had. This allowed ABCs and HDCs to be on the same reference frame and to have a one-to-one correspondence on preferred head directions. A single SBC was connected to all ABCs with corresponding HDCs' synapses in the dendritic trees. A single ABC fired with maximum activity only if both its corresponding HDC and related SBC were activated.

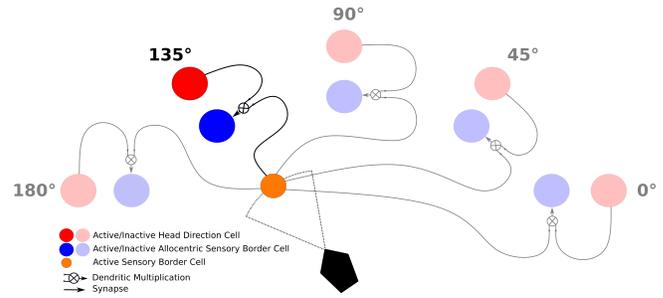

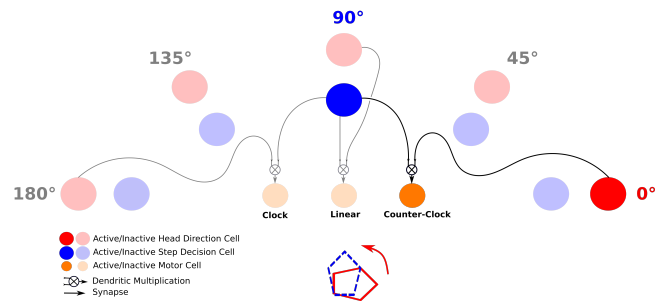

**Figure 2: Dendritic multiplication examples. (a) Transforming egocentric (camera-reference) depth information to allocentric (world-reference) information of a detected obstacle; (b) Generating egocentric motor commands (here, turning by 90 degrees to the left).**

## 3.3 Place Cells exhibiting STDP

In our model, PCs represented the robot's location by firing maximally when the robot was on the neuron's preferred place field. To computationally represent the PC activity, we integrated synaptic inputs from different layers of GCs, following theoretical studies [25, 61]. One of the advantage of GC coding is the ability to represent an environment of any size using a CAN model [21]. The GC-PC circuit approach that we incorporated in our network integrated this advantage into PC coding.

Interestingly enough, experiments on rats give evidence that the preferred place fields of PCs remap every time a rat enters a new environment, while the preferred place fields of GCs remain stable [20]. We employed this experimental finding in our model by not allowing any pre-defined (i.e., static) synaptic connection between GCs and PCs. Specifically, we regarded the remapping of the PCs' preferred place fields as a reassignment of the limited neural resources in the hippocampus, as follows. Prior to the robot entering a new environment, there were no synapses between rectangular GCs and PCs. Upon GC and PC concurrent (within tens of milliseconds) activation, new synapses were generated through STDP. Since there was no input to PCs from rectangular GCs before generating any synapse, an external input was initially used to drive the PCs.

To overcome the practical limitation of recruiting PCs representing all combinations of 3 layers of GCs, which was more than $10^6$ possible locations in the simulated environment, we used only a small number of locations to represent the borders of the environment. Therefore, we used instead only 500 neurons that were allocated as PCs before the robot entered a new environment. Following a complete exploration of an environment, half of the PCs became border sensitive place cells (PC0), having place fields near the borders and the other half of the PCs became general place cells (PC1) and had larger place fields at arbitrary locations.

PCs were activated by the external inputs in a sequential and exclusive manner as follows. Every PC had global inhibitory synaptic connections to all other PCs. When a PC was strongly activated, then all other PCs were inhibited and none of them would respond to external inputs. If there was no PC activated, a PC that had not been recruited before would respond to an external input. This PC turned into PC0 if the external input came from SBCs. Otherwise, this PC turned into PC1. This way, we were able to have a neural representation of the environment employing only a limited number of PCs with different preferred place field locations and different place field sizes (Figure 3B).

## 3.4 A Distributed Map of the Environment

The environment was memorized in a distributed fashion, using the synaptic weights between PCs in the hippocampus and BCs in the subiculum. In other words, there was no global map stored in a single location in the model. In addition, upon activation of the PCs with preferred place fields associated with specific locations, the nearby BCs were also activated to provide a partial map of the environment surrounding the location. This partial map provided by BCs gave the robot positions of the nearby borders and impeded it from hitting into the walls.

One of the advantages of such distributed learning is that it requires a much smaller number of neurons than what a global map would need, by not keeping the topological structure of the environment. This allowed us to keep the limited neural resources for other navigation functions, including route planning or localization.

Goal cells were only activated at locations near distinct goals. Here we defined 4 different goals represented by 4 different colors on the wall at different locations. Goal Detection Cells (GDC) fired when a wall painted with one of the goal colors was observed in the RGB-Depth camera. Subsequently, the sensory border information was integrated into GDC outputs to determine whether the goal was close to the robot, or not.

Our model employed STDP to strengthen synapses with different levels of weights between PCs and BCs. BCs first got inputs from ABCs when no synapses were generated from PCs. The weight of the synapse between PC and BC was updated when both neurons fired within tens of milliseconds.

## 3.5 Motor Commands to the Robot

Figure 2b shows the transformation from the robot's desired moving direction presented by Step Decision Cells (SDCs) to motor commands decoded from Motor Cells (MCs). In our model, there were 8 SDCs representing 8 equally spaced desired moving directions (at 0, 45, 90, 135, 180, 225, 270, 315 degrees respectively) for the robot. There were 3 MCs, one was a Linear velocity Motor Cell (LMC) representing linear velocity and the other two were Angular velocity Motor Cells (AMCs) representing angular velocity of clockwise and counter-clockwise rotations, respectively. Motor commands sent to the robot were decoded based on linear transformations performed on MCs' neural activities.

Emulating a typical walking pattern found in primates, the robot could move only in the direction that it was facing, depicted by the head direction. We created dendrites in MCs that received synaptic inputs from SDCs. For computational efficiency, HDC inputs were down-sampled from 360 to 8 head directions corresponding to the 8 preferred directions of SDCs. Through dendritic multiplication, LMC was only activated when the HDC and SDC inputs shared the same preferred direction. In that way, the robot could move forward only if it was heading the desired moving direction.

When the robot was not facing the desired moving direction, 2 AMCs were competing with each other to encode the angular velocity and the rotation direction. Each AMC received excitatory inputs from HDCs and SDCs; It also received and gave inhibitory inputs from and to the other AMCs. The down-sampled HDC and the SDC inputs formed dendritic trees with dendritic multiplications on AMCs, the tree connections were formed based on the shortest rotation distance principle described below. Whenever the robot needed to rotate, the shortest rotation distance principle dictated the robot to rotate in the direction with shortest distance to the desired moving direction. For smoother movements, HDC inputs for AMCs were also down-sampled from 360 to 36 head directions.

Figure 2b shows an example on how the dendritic trees were formed. Let us assume that the robot had a desired moving direction represented by an SDC with preferred direction of 90 degrees and a head direction represented by a down-sampled HDC with preferred direction 0 degrees. Clearly, the robot should rotate in the counter-clockwise direction based on the shortest rotation distance principle. Therefore, these 2 activated neurons would connect to the AMC representing counter-clockwise direction to form a dendritic tree. Through dendritic multiplication, the AMC representing counter-clockwise direction had stronger activity than the AMC representing clockwise direction. The latter AMC was inhibited, and the robot turned counter-clockwise. Although here we limit the step decisions to 8 directions, our model can generalize to any number of possible directions, as long as it is not larger than the number of HDCs.

# 4 COMPUTATIONAL NEURAL ELEMENTS

## 4.1 Spiking Neurons

Spiking neurons were simulated in real-time as LIF models [31].

$$\tau_m \frac{dv}{dt} = -v + rI_{sum} \tag{1}$$

where $v$ is the membrane voltage of neuron at time $t$, $I_{sum}$ is the summation input currents from all synaptic connections, $\tau_m$ is the membrane time constant and $r$ is the membrane resistance.

## 4.2 Synapse and Synaptic Plasticity

To allow for real-time computation in software, where saving and computing a list of spike times is inefficient, we introduced a trace model that tracked pre-synaptic neural activities with either fixed or plastic synaptic weights. We updated the spike trace with incoming spikes from pre-synaptic neuron, as follows:

$$\frac{dI}{dt} = -\frac{I}{\tau} + \sum_s \delta(t - t^s) \tag{2}$$

where $I$ is the spike trace in the synapse at time $t$,$\tau$ is the decay factor controls how fast the synapse forgets, and $t^s$ is the time of the pre-synaptic spike.

Plastic synaptic weights were changed based on a pair-based model of Hebbian STDP [51]. The model assumes that synaptic weight raises when pre-synaptic neuron generates a spike in a very short range of time (tens of milliseconds) ahead of a spike generated by post-synaptic neuron. We implemented Long-term Potentiation (LTP) by using a spike trace $x$ to keep track of the pre-synaptic neuron's spikes. When pre-synaptic neuron $j$ spiked, trace $x_j$ was updated by one, otherwise $x_j$ decayed fast. When post-synaptic neuron $i$ spiked, weight $w_{ij}$ was updated base on the value of $x_j$. This is formalized below:

$$\frac{dx_j}{dt} = -\frac{x_j}{\tau} + \sum_s \delta(t - t_j^s)$$

$$\frac{dw_{ij}}{dt} = Amp(w_{ij})x_j \sum_s \delta(t - t_i^s) \tag{3}$$

$$Amp(w) = A(w_{max} - w)$$

where $x_j$ is the spike trace for pre-synaptic neuron $j$, $\tau$ is the decay factor controls the time window size of STDP, $t_j^s$ is the time of the pre-synaptic spike, $t_i^s$ is the time of the post-synaptic spike, $w_{ij}$ is the weight between pre-synaptic neuron $j$ and post-synaptic neurons $i$, $Amp(w)$ controls the weight changing amplitude and maximum weight $w_{max}$, $A$ is the amplitude factor.

## 4.3 Nonlinear and Linear Dendritic Computations

Neural information underwent a nonlinear transformation using dendritic multiplication of synaptic inputs. This allowed the dendritic tree to exhibit a strong current input towards the neuron only if both synapses were strongly activated (Figure 2). This was used in various layers at the network to integrate information between neural layers, e.g., between HDC and ABC, to form a representation for the allocentric border information from ABC, or between SBC and GDC, to represent goal information. To efficiently compute dendritic multiplication on large groups of neurons, we defined a matrix representation. Allowing no more than 2 synapses per dendritic tree, the matrix is as follows:

$$D_{multiply} = \begin{bmatrix} A_{11} & \dots & A_{1m} \\ \dots & \dots & \dots \\ A_{n1} & \dots & A_{nm} \end{bmatrix} \begin{bmatrix} B_1 \\ \dots \\ B_m \end{bmatrix} \tag{4}$$

where $n$ was the number of post-synaptic neurons with $m$ dendritic trees for each neuron, $A$ and $B$ are the two synapses contained on each dendritic tree. Here, synapses $B$ were represented as a vector because we assumed that for a given dendritic tree, the same synapses $B$ correspond to different post-synaptic neurons. On the contrary, synapses $A$ could be different for different neurons, and therefore, synapses $A$ for dendritic trees of a single neuron were represented as columns in the matrix. This allowed us to compute the integrated dendritic inputs for n neurons using a single matrix-vector multiplication.

Each neuron received current inputs from multiple dendritic trees, and the overall current input for a single LIF neuron in equation (1) was the summation of inputs from all dendritic trees. Each dendritic tree contained arbitrary number of synapses. We allowed for one of the two kinds of dendritic computations, summation or multiplication, to be used on each dendritic tree.

# 5 EXPERIMENTAL ENVIRONMENT

We implemented our SNN in the Robotic Operating System (ROS) environment in a distributed and modular framework [57], which is very similar to that of a real mammalian brain. Specifically, ROS nodes were packaged into separate threads and computed using parallel multithreads. There were 4 kinds of nodes in our ROS based neural system, representing neurons, synapses, plastic synapses, and dendrites. Nodes communicated using messages wrapped in topics, in a similar fashion that neurons communicate by sending packages of chemical information in synapses. These nodes communicated using 2 kinds of messages, spikes and current. A neuron node published spike messages to synapse nodes, and a synapse node published current messages to neuron nodes. To allow for a real time spiking system, nodes were updated every 10 milliseconds. To run the experiment, we used the Gazebo simulator [36].

# 6 RESULTS

## 6.1 SNN-controlled Robot

The movement of the robot was autonomous. The control decision signal was the only external input to the robot used to provide heuristic decision strategies in different control modes. To efficiently explore the environment and evaluate the SNN's ability to learn its surroundings, we designed a task having 3 stages. First, the robot followed the walls of the environment for 30 minutes; This procedure required the recruitment of a number of specialized cells, e.g., BCs, PCs, and GCs. Through this procedure, the robot followed the walls and generated a map of the double T-maze efficiently without entering the inner area of the maze. Second, the robot explored the environment randomly; This procedure recruited more PCs as the robot did a random search of the environment without colliding with the learned borders. Third, visual inputs were turned off and the robot walked through the learned environment, for more than 2 hours, without hitting the memorized walls. The proposed SNN controlled the robot working in an end-to-end manner. It received sensory inputs from the robot's sensors and generated motor commands to the robot. At each time step, a robotic control command was decoded from MCs.

## 6.2 Place Cell Generation

Each neuron in the PC layer turned into PC0 (red circles; Figure 3B) or PC1 (blue circles; Figure 3B) upon creating synapse with the GCs. When following the wall, the robot moved parallel to the wall if it stayed in a place field of an activated PC0. When there was no PC0 activated, GCs received input from external control decision and the SNN drove the robot to turn towards the border to recruit a new PC0. A PC1 was generated when the robot entered a location where no PC was activated.

Upon learning the environment, PC0s and PC1s were generated and had place fields sampling the possible locations that the robot can reach. During the 2.5 hour experiment, 182 PC0s and 60 PC1s were generated, the latter to support future route planning functions. Most of the PC0s were generated during the robot following the border. In total, our model only used 242 PCs to represent the environment, which is much less than the upper boundary of $10^6$ GCs combinations.

## 6.3 Learning obstacles and goals

When a PC0 was generated, it also served as an anchor neuron for episodic memories within or surrounding its place field. Interestingly, the hippocampus has been associated with episodic memories since the famous research on patient HM [62]. Our model could encode two types of episodic memories, the location of environmental borders and that of different goals.

In Figure 3 we show the spiking activity of 2 BCs and 4 GCs during learning the environment, and after memorizing it (i.e., turning off the visual input). In Figure 3D, red dots represent the conditional firing of BCs that are encoding a south border; In Figure 2F, the colored dots represent the activity of one of the 4 GoalCs when the robot observed the goal.

BCs and GoalCs could be activated by visual inputs. When vision was turned off, the neural representation of the environment was represented in the learned BCs and GoalCs. These results also align with experimental studies on rats, being able to orient themselves in the dark [20].

## 7 DISCUSSION

Here, we presented an SNN that emulates a neurophysiologically plausible connectome among specialized neural cells that have been associated with the brain's navigational system. By having most of its connections predefined by the brain's structure, our model required minimum learning to become functional in mapping the environment by itself. We showed how the model can be used as a robotic controller to create a map of an uncharted territory and learn the location of different goals. These results align with our overarching goal which is to develop an end-to-end neuromorphic controller for an autonomously moving robot that explores and reacts efficiently with its environment, much like primates do. In real-world, robots ask for real-time processing of fast-varying, noisy

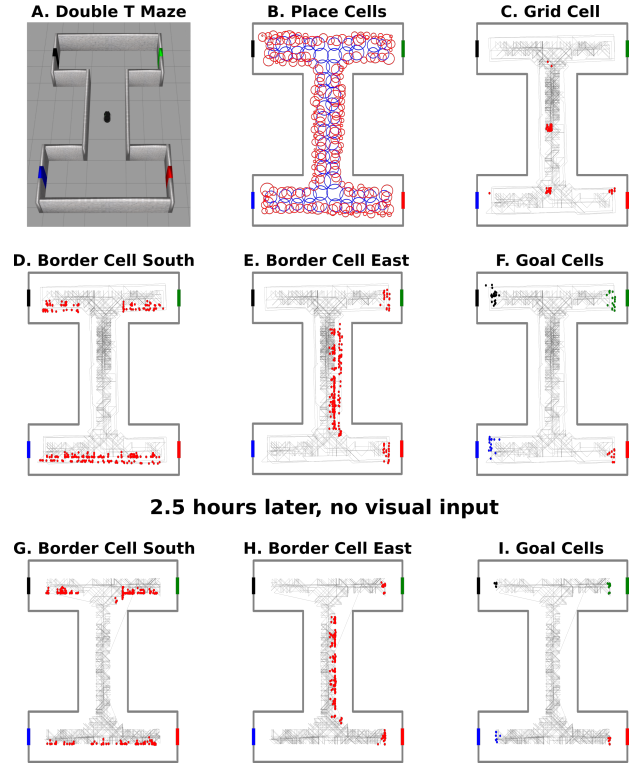

**Figure 3: Experimental results in a double T maze environment. Representative neural cells are shown, each dot represents the associated receptive field of the specialized neuron activated when the robot is at the specific location. B-F, cell activities during learning the environment. G-I, cell activities after learning (no visual input)**

information in a fast-changing environment, where the current convenient assumptions, such as deep learning algorithms devouring high volumes of ideally non-varying data, are hard to survive.

To be effective in real-world, an autonomous agent, either biological or artificial, should 1) be robust to a noisy neural representation, 2) adapt to a fast-changing environment, 3) learn with no or limited supervision or reinforcement, and 4) compute efficiently with resource limitations. Our biologically constrained SNN overcomes the main problem that SNN architectures have, that of slow or computationally inefficient learning, and paves the way for introducing non-neuronal cells, such as astrocytes, that also process and learn information in the brain [38, 39, 55]. The application of SNNs in controlling a behavior, such as a motor task has indeed been impeded mainly by the lack of efficient or biologically-constrained learning methods [5, 13, 15, 64]. Contrary to typical neuron models in conventional multi-layer networks that can be optimized to perform complex computational tasks [41], spiking neurons are typically limited to local learning rules and usually lack well-defined objective functions. To introduce functionality into an SNN, neural spikes are constrained at desired times through engineering [6, 22, 47, 54] or biologically-inspired [17, 32, 43, 56] methods; these direct approaches, however, narrow the network's scale and, subsequently, range of applicability. SNNs can become more flexible through indirect methods that first find a solution via conventional networks and then replicate it with SNNs [1, 27]. Recent applications of gradient-descent alternatives to SNNs [26, 42] are promising but, lacking biological relevance, can only match the network's input to its output, much regarding the network as a black-box.

In fact, an additional benefit from developing neural-controlled robots is that they can serve as testbeds to inform brain scientists on how the neural system could be structured to function properly. Evolved from typical multi-layer networks, SNNs increase the level of realism in neural networks by replacing the arbitrary input-output function of neurons with a set of differential equations that describe the electrical charge of the neuron's membrane [24, 30, 40]. Imitating their biological counterparts, neurons in the SNN component fire not on each propagation cycle, as in typical networks [19, 41], but asynchronously to each other. However, this asynchronous character of the SNNs gives a synchronous rise to their main strength ,biological realism, and weakness: SNNs are computationally inefficient on traditional Von Neumann architecture [1].

Recent developments on large-scale neuromorphic hardware [12, 18, 48, 59] makes SNNs ideal for serving non-Von Neumann architectures with asynchronous design implementations. Without a globally distributed clock for data synchronization, neuromorphic hardware computes locally, using a limited number of cores. This makes the device more power and computationally efficient than traditional Von Neumann architecture. Therefore, SNNs, with carefully designed biological constrains, show a great potential to be computationally efficient on neuromorphic hardware. Similar to the SNN model we proposed in this paper, an SNN that can seamlessly be run in a neuromorphic hardware should 1) learn synaptic weight based on spike information from a limited number of neurons, 2) control, maintain and change the information flow using structured synaptic connections rather than all-to-all connections, and 3) activate a small number of neurons at any given time. Our approach that brings knowledge representation and computation down to the neural level shows a great potential to realize the promises for robustness, adaptability and efficiency, when engrafted into a neuromorphic hardware.

## 8 CONCLUSION

Studies on biologically realistic networks exhibiting a particular behavior are of utmost importance in advancing Brain Science and Robotics, two fast growing fields that are also converging via biomimicry and artificial intelligence. Bringing brain-mimesis further down the cellular level, the built-in learning mechanisms of an SNN model of the biological spatial system, show a potential for being unleashed into intrinsically adaptive robotic controllers that will also help further our understanding of brain function and dysfunction. This will catalyze efforts towards augmenting human cognitive abilities and engrafting them into intelligent autonomous robots.

## REFERENCES
[1] LF Abbott, Brian DePasquale, and Raoul-Martin Memmesheimer. 2016. Building functional networks of spiking model neurons. *Nature neuroscience* 19, 3 (2016),

350.
[2] Eric A Antonelo, Benjamin Schrauwen, Xavier Dutoit, Dirk Stroobandt, and Marnix Nuttin. 2007. Event detection and localization in mobile robot navigation using reservoir computing. In *International Conference on Artificial Neural Networks*. Springer, 660–669.
[3] Caswell Barry and Neil Burgess. 2014. Neural mechanisms of self-location. *Current Biology* 24, 8 (2014), R330–R339.
[4] Danielle S Bassett and Olaf Sporns. 2017. Network neuroscience. *Nature neuroscience* 20, 3 (2017), 353.
[5] Martin Boerlin, Christian K Machens, and Sophie Denève. 2013. Predictive coding of dynamical variables in balanced spiking networks. *PLoS computational biology* 9, 11 (2013), e1003258.
[6] Sander M Bohte, Joost N Kok, and Han La Poutre. 2002. Error-backpropagation in temporally encoded networks of spiking neurons. *Neurocomputing* 48, 1-4 (2002), 17–37.
[7] Yoram Burak and Ila R Fiete. 2009. Accurate path integration in continuous attractor network models of grid cells. *PLoS computational biology* 5, 2 (2009), e1000291.
[8] Zhiqiang Cao, Long Cheng, Chao Zhou, Nong Gu, Xu Wang, and Min Tan. 2015. Spiking neural network-based target tracking control for autonomous mobile robots. *Neural Computing and Applications* 26, 8 (2015), 1839–1847.
[9] Taylor S Clawson, Silvia Ferrari, Sawyer B Fuller, and Robert J Wood. 2016. Spiking neural network (SNN) control of a flapping insect-scale robot. In *Decision and Control (CDC), 2016 IEEE 55th Conference on*. IEEE, 3381–3388.
[10] Taylor S Clawson, Terrence C Stewart, Chris Eliasmith, and Silvia Ferrari. 2017. An adaptive spiking neural controller for flapping insect-scale robots. In *Computational Intelligence (SSCI), 2017 IEEE Symposium Series on*. IEEE, 1–7.
[11] Sakyasingha Dasgupta, Florentin Wörgötter, and Poramate Manoonpong. 2013. Information dynamics based self-adaptive reservoir for delay temporal memory tasks. *Evolving Systems* 4, 4 (2013), 235–249.
[12] Mike Davies, Narayan Srinivasa, Tsung-Han Lin, Gautham Chinya, Prasad Joshi, Andrew Lines, Andreas Wild, and Hong Wang. 2018. Loihi: A Neuromorphic Manycore Processor with On-Chip Learning. *IEEE Micro* (2018).
[13] Brian DePasquale, Mark M Churchland, and LF Abbott. 2016. Using firing-rate dynamics to train recurrent networks of spiking model neurons. *arXiv preprint arXiv:1601.07620* (2016).
[14] Travis DeWolf, Terrence C Stewart, Jean-Jacques Slotine, and Chris Eliasmith. 2016. A spiking neural model of adaptive arm control. *Proc. R. Soc. B* 283, 1843 (2016), 20162134.
[15] Chris Eliasmith. 2005. A unified approach to building and controlling spiking attractor networks. *Neural computation* 17, 6 (2005), 1276–1314.
[16] Uğur M Erdem and Michael Hasselmo. 2012. A goal-directed spatial navigation model using forward trajectory planning based on grid cells. *European Journal of Neuroscience* 35, 6 (2012), 916–931.
[17] Răzvan V Florian. 2007. Reinforcement learning through modulation of spike-timing-dependent synaptic plasticity. *Neural Computation* 19, 6 (2007), 1468–1502.
[18] Steve B Furber, Francesco Galluppi, Steve Temple, and Luis A Plana. 2014. The spinnaker project. *Proc. IEEE* 102, 5 (2014), 652–665.
[19] Ian Goodfellow, Yoshua Bengio, Aaron Courville, and Yoshua Bengio. 2016. *Deep learning*. Vol. 1. MIT press Cambridge.
[20] Roddy M Grieves and Kate J Jeffery. 2017. The representation of space in the brain. *Behavioural processes* 135 (2017), 113–131.
[21] Alexis Guanella, Daniel Kiper, and Paul Verschure. 2007. A model of grid cells based on a twisted torus topology. *International journal of neural systems* 17, 04 (2007), 231–240.
[22] Robert Gütig and Haim Sompolinsky. 2006. The tempotron: a neuron that learns spike timing–based decisions. *Nature neuroscience* 9, 3 (2006), 420.
[23] Elizabeth JO Hamel, Benjamin F Grewe, Jones G Parker, and Mark J Schnitzer. 2015. Cellular level brain imaging in behaving mammals: an engineering approach. *Neuron* 86, 1 (2015), 140–159.
[24] Alan L Hodgkin and Andrew F Huxley. 1952. A quantitative description of membrane current and its application to conduction and excitation in nerve. *The Journal of physiology* 117, 4 (1952), 500–544.
[25] Guanwen Huang, Bailu Si, and Fengzhen Tang. 2017. Model learning based on grid cell representations. In *Robotics and Biomimetics (ROBIO), 2017 IEEE International Conference on*. IEEE, 1032–1037.
[26] Dongsung Huh and Terrence J Sejnowski. 2017. Gradient descent for spiking neural networks. *arXiv preprint arXiv:1706.04698* (2017).
[27] Eric Hunsberger and Chris Eliasmith. 2015. Spiking deep networks with LIF neurons. *arXiv preprint arXiv:1510.08829* (2015).
[28] Tiffany Hwu, Jeffrey Krichmar, and Xinyun Zou. 2017. A complete neuromorphic solution to outdoor navigation and path planning. In *Circuits and Systems (ISCAS), 2017 IEEE International Symposium on*. IEEE, 1–4.
[29] Tiffany Hwu, Alexander Wang, Nicolas Oros, and Jeffrey Krichmar. 2017. Adaptive robot path planning using a spiking neuron algorithm with axonal delays. *IEEE Transactions on Cognitive and Developmental Systems* (2017).

[30] Eugene M Izhikevich. 2003. Simple model of spiking neurons. *IEEE Transactions on neural networks* 14, 6 (2003), 1569–1572.

[31] Eugene M Izhikevich. 2004. Which model to use for cortical spiking neurons? *IEEE transactions on neural networks* 15, 5 (2004), 1063–1070.

[32] Eugene M Izhikevich. 2007. Solving the distal reward problem through linkage of STDP and dopamine signaling. *Cerebral cortex* 17, 10 (2007), 2443–2452.

[33] Eduardo J Izquierdo and Randall D Beer. 2013. Connecting a connectome to behavior: an ensemble of neuroanatomical models of C. elegans klinotaxis. *PLoS computational biology* 9, 2 (2013), e1002890.

[34] Travis A Jarrell, Yi Wang, Adam E Bloniarz, Christopher A Brittin, Meng Xu, J Nichol Thomson, Donna G Albertson, David H Hall, and Scott W Emmons. 2012. The connectome of a decision-making neural network. *Science* 337, 6093 (2012), 437–444.

[35] Philipp J Keller and Misha B Ahrens. 2015. Visualizing whole-brain activity and development at the single-cell level using light-sheet microscopy. *Neuron* 85, 3 (2015), 462–483.

[36] Nathan Koenig and Andrew Howard. 2004. Design and use paradigms for gazebo, an open-source multi-robot simulator. In *Intelligent Robots and Systems, 2004.(IROS 2004). Proceedings. 2004 IEEE/RSJ International Conference on*, Vol. 3. IEEE, 2149–2154.

[37] Nancy J Kopell, Howard J Gritton, Miles A Whittington, and Mark A Kramer. 2014. Beyond the connectome: the dynome. *Neuron* 83, 6 (2014), 1319–1328.

[38] Leo Kozachkov and Konstantinos P Michmizos. 2017. The Causal Role of Astrocytes in Slow-Wave Rhythmogenesis: A Computational Modelling Study. *arXiv preprint arXiv:1702.03993* (2017).

[39] Leo Kozachkov and Konstantinos P Michmizos. 2017. A Computational Role for Astrocytes in Memory. *arXiv preprint arXiv:1707.05649* (2017).

[40] Louis Lapicque. 1907. Recherches quantitatives sur l'excitation electrique des nerfs traitee comme une polarization. *Journal de physiologie et de pathologie générale* 9 (1907), 620–635.

[41] Yann LeCun, Yoshua Bengio, and Geoffrey Hinton. 2015. Deep learning. *nature* 521, 7553 (2015), 436.

[42] Jun Haeng Lee, Tobi Delbruck, and Michael Pfeiffer. 2016. Training deep spiking neural networks using backpropagation. *Frontiers in neuroscience* 10 (2016), 508.

[43] Robert Legenstein, Dejan Pecevski, and Wolfgang Maass. 2008. A learning theory for reward-modulated spike-timing-dependent plasticity with application to biofeedback. *PLoS computational biology* 4, 10 (2008), e1000180.

[44] Jeff W Lichtman and Winfried Denk. 2011. The big and the small: challenges of imaging the brain's circuits. *Science* 334, 6056 (2011), 618–623.

[45] Chit-Kwan Lin, Andreas Wild, Gautham N Chinya, Tsung-Han Lin, Mike Davies, and Hong Wang. 2018. Mapping spiking neural networks onto a manycore neuromorphic architecture. In *Proceedings of the 39th ACM SIGPLAN Conference on Programming Language Design and Implementation*. ACM, 78–89.

[46] Michael London and Michael Häusser. 2005. Dendritic computation. *Annu. Rev. Neurosci.* 28 (2005), 503–532.

[47] Raoul-Martin Memmesheimer, Ran Rubin, Bence P Ölveczky, and Haim Sompolinsky. 2014. Learning precisely timed spikes. *Neuron* 82, 4 (2014), 925–938.

[48] Paul A Merolla, John V Arthur, Rodrigo Alvarez-Icaza, Andrew S Cassidy, Jun Sawada, Filipp Akopyan, Bryan L Jackson, Nabil Imam, Chen Guo, Yutaka Nakamura, et al. 2014. A million spiking-neuron integrated circuit with a scalable communication network and interface. *Science* 345, 6197 (2014), 668–673.

[49] Michael J Milford, Janet Wiles, and Gordon F Wyeth. 2010. Solving navigational uncertainty using grid cells on robots. *PLoS computational biology* 6, 11 (2010), e1000995.

[50] Michael J Milford, Gordon F Wyeth, and David Prasser. 2004. RatSLAM: a hippocampal model for simultaneous localization and mapping. In *Robotics and Automation, 2004. Proceedings. ICRA'04. 2004 IEEE International Conference on*, Vol. 1. IEEE, 403–408.

[51] Abigail Morrison, Markus Diesmann, and Wulfram Gerstner. 2008. Phenomenological models of synaptic plasticity based on spike timing. *Biological cybernetics* 98, 6 (2008), 459–478.

[52] Edvard I Moser, Emilio Kropff, and May-Britt Moser. 2008. Place cells, grid cells, and the brain's spatial representation system. *Annual review of neuroscience* 31 (2008).

[53] Stefano Nolfi, Josh C Bongard, Phil Husbands, and Dario Floreano. 2016. Evolutionary Robotics.

[54] Jean-Pascal Pfister, Taro Toyoizumi, David Barber, and Wulfram Gerstner. 2006. Optimal spike-timing-dependent plasticity for precise action potential firing in supervised learning. *Neural computation* 18, 6 (2006), 1318–1348.

[55] Ioannis Polykretis, Vladimir Ivanov, and Konstantinos P. Michmizos. 2018. A Neural-Astrocytic Network Architecture: Astrocytic calcium waves modulate synchronous neuronal activity. *ICONS '18: International Conference on Neuromorphic Systems, July 23–26, 2018, Knoxville, TN, USA* (2018). https://doi.org/10.1145/3229884.3229890

[56] Filip Ponulak and Andrzej Kasiński. 2010. Supervised learning in spiking neural networks with ReSuMe: sequence learning, classification, and spike shifting. *Neural computation* 22, 2 (2010), 467–510.

[57] Morgan Quigley, Ken Conley, Brian Gerkey, Josh Faust, Tully Foote, Jeremy Leibs, Rob Wheeler, and Andrew Y Ng. 2009. ROS: an open-source Robot Operating System. In *ICRA workshop on open source software*, Vol. 3. Kobe, Japan, 5.

[58] Yadira Quiñonez, Mario Ramirez, Carmen Lizarraga, Iván Tostado, and Juan Bekios. 2015. Autonomous robot navigation based on pattern recognition techniques and artificial neural networks. In *International Work-Conference on the Interplay Between Natural and Artificial Computation*. Springer, 320–329.

[59] Johannes Schemmel, Daniel Brüderle, Andreas Grübl, Matthias Hock, Karlheinz Meier, and Sebastian Millner. 2010. A wafer-scale neuromorphic hardware system for large-scale neural modeling. In *Circuits and systems (ISCAS), proceedings of 2010 IEEE international symposium on*. IEEE, 1947–1950.

[60] Stefan Schliebs and Nikola Kasabov. 2013. Evolving spiking neural networkâĂŤa survey. *Evolving Systems* 4, 2 (2013), 87–98.

[61] Trygve Solstad, Edvard I Moser, and Gaute T Einevoll. 2006. From grid cells to place cells: a mathematical model. *Hippocampus* 16, 12 (2006), 1026–1031.

[62] Larry R Squire and Stuart Zola-Morgan. 1991. The medial temporal lobe memory system. *Science* 253, 5026 (1991), 1380–1386.

[63] Jeffrey S Taube. 2007. The head direction signal: origins and sensory-motor integration. *Annu. Rev. Neurosci.* 30 (2007), 181–207.

[64] Dominik Thalmeier, Marvin Uhlmann, Hilbert J Kappen, and Raoul-Martin Memmesheimer. 2016. Learning universal computations with spikes. *PLoS computational biology* 12, 6 (2016), e1004895.

[65] Duncan J Watts and Steven H Strogatz. 1998. Collective dynamics of 'small-world' networks. *nature* 393, 6684 (1998), 440.

[66] John G White, Eileen Southgate, J Nichol Thomson, and Sydney Brenner. 1986. The structure of the nervous system of the nematode Caenorhabditis elegans. *Philos Trans R Soc Lond B Biol Sci* 314, 1165 (1986), 1–340.